\documentclass[11pt,a4paper]{article}
\usepackage{amsmath}
\usepackage{amssymb}
\usepackage{amsthm}
\usepackage{epsfig}
\usepackage{times}
\usepackage{euler}
\usepackage{wrapfig}

\usepackage[hang,stable]{footmisc}
\newcommand{\Scal}{{\cal S}}
\newcommand{\sss}{\scriptscriptstyle}
\newcommand{\reals}{\mathbb{R}}
\newcommand{\complex}{\mathbb{C}}
 
\newcommand{\diff}{\mathrm{Diff(M)}} 
\newcommand{\diffmap}{f} 
\newcommand{\stab}{\text{Stab}}
\begin{document}
\ifx\href\undefined\else\hypersetup{linktocpage=true}\fi

\newtheorem{theorem}{Theorem}
\theoremstyle{definition}
\newtheorem{definition}{Definition}

\title{Some remarks on the notions of general covariance\\
       and background independence}
\author{\normalsize Domenico Giulini            \\
        \normalsize University of Freiburg      \\
        \normalsize Department of Physics       \\
        \normalsize Hermann-Herder-Strasse 3    \\
        \normalsize D-79104 Freiburg,           \\
        \normalsize Germany}

\date{}

\maketitle

\begin{abstract}
\noindent
The notion of `general covariance' is intimately related to 
the notion of `background independence'. Sometimes these 
notions are even identified. Such an identification was made 
long ago by James Anderson, who suggested to \emph{define} 
`general covariance' as absence of what he calls `absolute 
structures', a term here taken to define the even less concrete 
notion of `background'. We discuss some of the well known 
difficulties that occur when one tries to give a precise 
definition of the notion of `absolute structure'. As a result, 
there still seem to be fundamental difficulties in defining 
`general covariance' or `background independence' so as to 
become a non-trivial selection principle for fundamental 
physical theories.  

In the second part of this contribution we make some historical 
remarks concerning  the 1913 `Entwurf'-Theory by Einstein and 
Grossmann, in which general covariance was first put to the fore, 
and in which Einstein presented an argument why Poincar\'e-invariant 
theories for a zero-mass scalar gravitational field 
necessarily suffer from severe inconsistencies concerning 
energy conservation. This argument is instructive, even though---or 
because---it appears to be incorrect, as we will argue below. 

This paper is a contribution to ``An assessment of current 
paradigms in the physics of fundamental interactions'', edited 
by I.O.~Stamatescu (Springer Verlag, to appear).

\end{abstract}
\newpage

\newpage
\setcounter{tocdepth}{3}
\tableofcontents
\newpage

\section{Introduction}
It is a widely shared opinion that \emph{the} most 
outstanding and characteristic feature of General Relativity 
is its manifest \emph{background independence}.  Accordingly, those 
pursuing the canonical quantization programme for General 
Relativity see the fundamental virtue of their approach in 
precisely this preservation of  `background independence'. 
Indeed, there is no disagreement as to the background dependence 
of competing approaches, like the perturbative spacetime 
approach\footnote{Usually referred to as the `covariant 
approach', since perturbative expansions are made around a 
maximally symmetric spacetime, like Minkowski or DeSitter 
spacetime, and the theory is intended to manifestly keep 
covariance under this symmetry group (i.e. the Poincar\'e 
or the DeSitter group), not the diffeomorphism group!} 
or string theory. Accordingly, many string theorists would 
subscribe to the following research strategy: 

\begin{quote}
``Seek to make progress by identifying the background structure 
in our theories and removing it, replacing it with relations 
which evolve subject to dynamical laws.'' 
(\cite{Smolin:2005} p.\,10).
\end{quote} 
But what  means do we have to reliably identify background 
structures? 

There is another widely shared opinion 
according to which the principle of \emph{general covariance} is 
devoid of any physical content. This was first forcefully 
argued for in 1917 by Erich Kretschmann~\cite{Kretschmann:1917} 
and almost immediately accepted by Einstein~\cite{CPAE} 
(Vol.\,7, Doc.\,38, p.\,39), who from then on seemed to have 
granted the principle of general covariance no more physical 
meaning then that of a formal heuristic concept.  

From this it appears that it would not be a good idea to define 
`background independence' via `general covariance', for this would 
not result in a physically meaningful selection principle that 
could effectively guide future research. What would be a better 
definition? `Diffeomorphism invariance' is the most often quoted 
candidate. What precisely is the difference between general 
covariance and diffeomorphism invariance, and does the latter really 
improve on the situation? These are the questions to be discussed here. 
For related and partially complementary discussions, that also give 
more historical details, we refer to \cite{Norton:1993b,Norton:2003}
and \cite{Barbour:2001} respectively. 

As a historical remark we recall that Einstein quite clearly 
distinguished between the \emph{principle of general relativity (PGR)}
on one hand, and the \emph{principle of general covariance (PGC)} on the 
other. He proposed that the formal PGC would imply (but not be 
equivalent to) the physical PGR. He therefore adopted the PGC
as a heuristic principle, guiding our search for physically 
relevant equations. But how can this ever work if Kretschmann 
is right and hence PGC devoid of any physical content? Well, 
what Kretschmann precisely said was that \emph{any} physical law 
can be rewritten in an equivalent but generally covariant form. 
Hence general covariance alone cannot rule out any physical law.
Einstein maintained that it did if one considers the aspect of 
`formal simplicity'. Only those expressions which are formally 
`simple' after having been written in a generally covariant form 
should be considered as candidates for physical laws. 
Einstein clearly felt the lack for any good definition of
formal `simplicity', hence he recommended to experience it by 
comparing General Relativity to a generally covariant formulation  
of Newtonian gravity (then not explicitly known to him), which 
was later given by Cartan~\cite{Cartan:1923,Cartan:1924} and 
Friedrichs~\cite{Friedrichs:1927} and which did not turn out to 
be outrageously complicated, though perhaps somewhat 
unnatural. In any case, one undeniably feels that this state of 
affairs is not optimal.

\section{Attempts to define general covariance and/or \\ background independence}
A serious attempt to clarify the situation was made by James Anderson
\cite{Anderson:1964a}\cite{Anderson:RelativityPhysics}, who introduced 
the notion of \emph{absolute structure} which here we propose to take 
synonymously with background independence. This attempt will 
be discussed in some detail below. Before doing this we need 
to clarify some other notions.  

\subsection{Laws of motion: covariance versus invariance}
We represent space-time by a tuple $(M,g)$, where $M$ is a 
four-dimensional infinitely differentiable manifold and $g$ a 
Lorentzian metric of signature $(+,-,-,-)$. The global topology 
of $M$ is not restricted a priori, but for definiteness we shall 
assume a product-topology $\mathbb{R}\times S$ and think of the 
first factor as time and the second as space (meaning that $g$ 
restricted to the tangent spaces of the submanifolds 
$S_t:=\{t\}\times S$ is negative definite and positive definite 
along $\mathbb{R}_p:=\mathbb{R}\times\{p\}$. Also, unless stated 
otherwise, the Lorentzian metric $g$ is assumed to be at least 
twice continuously differentiable. We will generally not need to 
assume $(M,g)$ to be geodesically complete. 

Being a $C^\infty$-manifold, $M$ is endowed with a maximal atlas 
of coordinate functions on open domains in $M$ with $C^\infty$-transition 
functions on their mutual overlaps. Transition functions relabel the 
points that constitute $M$, which for the time being we think of as
recognizable entities, as mathematicians do. (For physicists these
points are mere `potential events' and do not have an obvious 
individuality beyond an actual, yet unknown, event that realizes this 
potentiality.) Different from maps between coordinate charts are global 
diffeomorphisms on $M$, which are $C^\infty$ maps $\diffmap:M\rightarrow M$
with $C^\infty$ inverses  $\diffmap^{-1}:M\rightarrow M$. 
Diffeomorphisms form a group (multiplication being composition) which 
we denote by $\diff$. Diffeomorphisms act (mostly, but not always, 
naturally) on geometric objects representing physical entities, like 
particles and fields.\footnote{For example, diffeomorphisms of $M$
lift naturally to any bundle associated to the bundle of linear 
frames and hence act naturally on spaces of sections in those 
bundles. In particular these include bundles of tensors of 
arbitrary ranks and density weights. On the other hand, there is no 
natural lift to e.g. spinor bundles, which are associated to the 
bundle of \emph{orthonormal} frames (which are only naturally acted upon 
by isometries, but not by arbitrary diffeomorphisms).} The transformed 
geometric object has then to be considered a priori as a 
\emph{different} object on the \emph{same} manifold (which is not meant 
to imply that they are necessarily physically distinguishable in a 
specific theoretical context). This is sometimes called the `active' 
interpretation of diffeomorphisms to which we will stick throughout.  

Structures that obey equations of motion are e.g. particles and fields. 
Classically, a structureless \emph{particle} (no spin etc.) is 
mathematically represented by a map \emph{into} spacetime: 
\begin{equation}
\label{eq:particle}
\gamma:\mathbb{R}\rightarrow M\,,  
\end{equation}
such that the tangent vector-field $\dot\gamma$ is everywhere timelike,
i.e. ($g(\dot\gamma,\dot\gamma)>0$). Other structures that are also 
represented by maps \emph{into} spacetime are strings, membranes, etc. 

A \emph{field} is defined by a map \emph{from} spacetime, that is,  
\begin{equation}
\label{eq:field}
\Phi: M\rightarrow V
\end{equation}
where $V$ is some vector space (or, slightly more general, affine 
space, to include connections). To keep the main argument simple 
we neglect more general situations where fields are sections in 
non-trivial vector bundles or non-linear target spaces.  

Let $\gamma$ collectively represent all structures given by maps 
into spacetime and $\Phi$ collectively all structures represented
by maps from spacetime. Equations of motions usually take the general 
symbolic form
\begin{equation}
\label{eq:eqmotions}
\mathcal{F}[\gamma,\Phi,\Sigma]=0
\end{equation}
which should be read as equation for $\gamma,\Phi$ \emph{given} 
$\Sigma$. 

$\Sigma$ represents some \emph{non-dynamical} structures on $M$.
Only if the value of $\Sigma$ is prescribed do we 
have definite equations of motions for $(\gamma,\Phi)$. This is 
usually how equations of motions are presented in physics: solve 
(\ref{eq:eqmotions}) for $(\gamma,\Phi$), \emph{given} $\Sigma$. 
Here only $(\gamma,\Phi)$ represent physical `degrees of freedom' 
of the theory to which alone observables refer (or out of which 
observables are to be constructed). By `theory' we shall always 
understand, amongst other things, a definite specification of 
degrees of freedom and observables.

The group $\diff$ acts on the objects $(\gamma,\Phi)$ 
(here we restrict the fields to tensor fields for simplicity) 
as follows:
\begin{subequations}
\label{eq:diff-action}
\begin{alignat}{3}
\label{eq:diff-action1}
(\diffmap,\gamma)\ \rightarrow\  & 
\diffmap\cdot\gamma\  :=\ && g\circ\gamma \qquad 
&& \mbox{for particles etc.}\,,\\
\label{eq:diff-action2}  
(\diffmap,\Phi)\ \rightarrow\  & 
\diffmap\cdot\Phi\  :=\  && D(\diffmap_*)\circ\Phi\circ\diffmap^{-1}\qquad 
&& \mbox{for fields etc.}\,,
\end{alignat}
\end{subequations}
where $D$ is the representation of $GL(4,\reals)$ carried by the fields. 
In addition, we require that the non-dynamical quantities $\Sigma$ to be 
geometric objects, i.e. to support an action of the diffeomorphism group.

\begin{definition}
\label{def:covariance}
Equation (\ref{eq:eqmotions}) is said to be \textbf{covariant} 
under the subgroup $G\subseteq\diff$ iff for all $\diffmap\in G$
\begin{equation}
\label{eq:def:covariance}
F[\gamma,\Phi,\Sigma]=0\,\Leftrightarrow\,
F[\diffmap\cdot\gamma\,,\,\diffmap\cdot\Phi\,,\,\diffmap\cdot\Sigma]=0\,.
\end{equation}
\end{definition}

\begin{definition}
\label{def:invariance}
Equation (\ref{eq:eqmotions}) is said to be \textbf{invariant} 
under the subgroup $G\subseteq\diff$ iff for all $\diffmap\in G$
\begin{equation}
\label{eq:def:invariance}
F[\gamma,\Phi,\Sigma]=0\,\Leftrightarrow\,
F[\diffmap\cdot\gamma\,,\,\diffmap\cdot\Phi\,,\,\Sigma]=0\,.
\end{equation}
\end{definition}
Note the difference: in Definition\,\ref{def:invariance} the 
non-dynamical structures $\Sigma$ are the same on both sides 
of the equation, whereas in Definition\,\ref{def:covariance} 
they are allowed to be also transformed by $\diffmap\in\diff$. 
Covariance merely requires the equation to  `live on the manifold', 
i.e. to be well defined in a differential-geometric sense, whereas 
an invariance is required to transforms solutions to the equations 
of motions to solutions of the \emph{very same} equation\footnote{
In the mathematical literature this is called a symmetry (of the 
equation). We wish to avoid the term `symmetry' here altogether because
that -- in our terminology -- is reserved for a further distinction of 
invariances into \emph{symmetries}, which change the physical state, 
and \emph{redundancies} (gauge transformations) which do not change the 
physical state. Here we will not need this distinction.}, which is 
a much more restrictive condition. 

As a simple example, consider the vacuum Maxwell equations on a 
fixed spacetime (Lorentzian manifold $(M,g)$):
\begin{subequations}
\label{eq:Maxwell}
\begin{alignat}{2}
\label{eq:Maxwell-1}
& dF\,&&=\,0\,,\\
\label{eq:Maxwell-2}
& d\star F\,&&=\,0\,,
\end{alignat}
\end{subequations}
where $F$ denotes the 2-form of the electromagnetic field and  
$d$ the exterior differential. The $\star$ denotes the (linear) 
`Hodge duality' map, which in components reads
\begin{equation}
\label{eq:def:Hodge-duality}
\star F_{\mu\nu}=\tfrac{1}{2}\varepsilon_{\mu\nu\alpha\beta}F^{\alpha\beta}\,,
\end{equation}
and which depends on the background metric $g$ through $\varepsilon$ 
and the operation of raising indices: 
$F^{\alpha\beta}:=g^{\alpha\mu}g^{\beta\nu}\,F_{\mu\nu}$. 
The system (\ref{eq:Maxwell}) is clearly $\diff$--covariant 
since it is written purely in terms of geometric structures on $M$
and makes perfect sense as equation on $M$. In particular, given any 
diffeomorphisms $\diffmap$ of $M$, we have that $\diffmap\cdot F$ satisfies 
(\ref{eq:Maxwell-1}) iff $F$ does. But it is \emph{not} likewise true 
that $d\star F=0$ implies $d\star\diffmap\cdot F=0$. In fact, it may be 
shown\footnote{This is true in 1+3 dimensions. In other dimensions 
higher than two  $\diffmap$ must even be an isometry of $g$.} that this 
is true iff $\diffmap$ is a conformal isometry of the background metric 
$g$, i.e. $\diffmap\cdot g=\lambda\, g$ for some positive real-valued 
function $\lambda$ on $M$. Hence the system 
(\ref{eq:Maxwell}) is not $\diff$--invariant
but only $G$--invariant, where $G$ is the conformal group of $(M,g)$.

\subsection{Triviality pursuit}
\subsubsection{Covariance trivialised (Kretschmann's point)}
Consider the ordinary `non-relativistic' diffusion equation for the 
$\reals$-valued field $\phi$ (giving the concentration density):
\begin{equation}
\label{eq:diffusion1}
\partial_t\phi=\kappa\Delta\phi\,.
\end{equation}
This does not look Lorentz covariant, let alone covariant under 
diffeomorphisms. But if rewritten it in the form 
\begin{equation}
\label{eq:diffusion2}
\left\{
n^\mu\nabla_\mu-\kappa(n^\mu n^\nu-
g^{\mu\nu})\nabla_\mu\nabla_\nu\right\}\phi=0\,,
\end{equation}
where $g^{\mu\nu}$ are the contravariant components of the spacetime 
metric (recall that we use the 'mostly minus' convention for its  
signature), $\nabla_\mu$ is its covariant derivative, and $n^\mu$ 
is a normalized covariant-constant timelike vector field which 
gives the preferred flow of time encoded in (\ref{eq:diffusion1})
(i.e. on scalar fields $\partial_t=n^\mu\nabla_\mu$). 
Equation (\ref{eq:diffusion2}) has the form (\ref{eq:eqmotions})
with no $\gamma$, $\Phi=\phi$, and $\Sigma=(g^{\mu\nu},n^\mu)$ and is 
certainly diffeomorphism covariant in the sense of 
Definition\,\ref{def:covariance}. The largest invariance group 
-- in the sense of Definition\,\ref{def:invariance} -- is given by 
that subgroup of $\diff$ whose elements stabilize the non-dynamical 
structures $\Sigma$. We write
\begin{equation}
\label{eq:StabilizerGroup}
\stab_\diff(\Sigma)=\{\diffmap\in\diff\mid \diffmap\cdot\Sigma=\Sigma\}
\end{equation}
In our case, $\stab_\diff(g)$ the 10-parameter Poincar\'e group. 
In addition, $\diffmap$ stabilizes $n^\mu$ if it is in the 7-parameter 
subgroup $\reals\times E(3)$ of time translations and spatial Euclidean
motions. 

This example already shows (there will be more below) how to proceed in
order to make any theory covariant under $\diff$. As already noted, 
$\diff$-covariance merely requires the equation to be well defined
in the sense of differential geometry, i.e. it should live on the 
manifold. It seems clear that any equation that has been written down 
in a special coordinate system on $M$ (like (\ref{eq:diffusion1})) 
can also be written in a $\diff$-covariant way by introducing the 
coordinate system -- or parts of it -- as background geometric 
structure. This is, in more modern terms, the formal core of the 
critique put forward by Erich Kretschmann in 1917~\cite{Kretschmann:1917}.

\subsubsection{Invariance trivialized}
Given that an equation of the form (\ref{eq:eqmotions}) is already 
$G$-covariant, we can equivalently express the condition of 
being $G$-invariant by
\begin{equation}
\label{eq:def:invariance2}
F[\gamma,\Phi,\Sigma]=0\,\Leftrightarrow\,
F[\gamma,\Phi\,,\,\diffmap\cdot\Sigma]=0\,,\quad\forall\diffmap\in G\,,  
\end{equation}
i.e. any solution of the equation parameterized by $\Sigma$ is also 
a solution of the \emph{different} equation parameterized by 
$\diffmap\cdot\Sigma$. Evidently, the more non-dynamical structures there 
are the more difficult it is to satisfy (\ref{eq:def:invariance2}). 
In generic situations it will only be satisfied if $G=\stab_\diff(\Sigma)$. 
Hence, in distinction to the covariance group, increasing the 
amount of structures of the type $\Sigma$ cannot enlarge the 
invariance group.  The case of the largest possible invariance 
group deserves a special name:   
\begin{definition}
\label{def:diff-invariance}
Equation (\ref{eq:eqmotions}) is called \textbf{diffeomorphism 
invariant} iff it allows $\diff$ as invariance group.
\end{definition}
In view of (\ref{eq:def:invariance2}), the requirement of 
$\diff$-invariance can be understood as a strong limit on the amount 
of non-dynamical structure $\Sigma$. Generically it seems to eliminate 
any $\Sigma$, i.e. the theory should contain no non-dynamical background 
fields whatsoever. Intuitively this is what background independence
stands for. Conversely, any $\diff$-covariant theory without 
non-dynamical fields is trivially $\diff$-invariant. Hence it seems 
sensible to simply identify `$\diff$-invariance' and `background 
independence', and this is what most working physicists seem to do. 

But this turns out to be too simple. The heart of the difficulty lies 
in our distinction between dynamical and non-dynamical structures, which 
turns out not to be sufficiently sharp. Basically we just said that 
a structure ($\gamma$ or $\Phi$) was dynamical if it had no a priori 
prescribed values, but rather obeyed some equations of motion. 
We did not say what qualifies an equation as an `equation of motion'. 
Can it just be \emph{any} equation?  If yes then we immediately object 
that there exists an obvious strategy to trivialize the requirement 
of $\diff$-invariance: just let the values of $\Sigma$ be determined 
by equations rather than by hand; in this way they formally become 
`dynamical' variables and no non-dynamical quantities are left. 
Formally this corresponds to the replacement 
scheme  
\begin{subequations}
\label{eq:trivialize}
\begin{alignat}{3}
\label{eq:trivialize1}
&\Phi   && \ \mapsto\ \Phi'\   &&=\ (\Phi,\Sigma)\,,\\
\label{eq:trivialize2}
&\Sigma && \ \mapsto\ \Sigma'\ &&=\ \emptyset    \,,
\end{alignat}
\end{subequations}
so that invariance now becomes as trivial as the requirement of 
covariance.

More concretely, reconsider the examples (\ref{eq:Maxwell}) 
and (\ref{eq:diffusion2}) above. In the first case we now regard 
the spacetime metric $g$ as `dynamical' field for which we add 
the condition of flatness as `equation of motion':
\begin{equation}
\label{eq:RiemZero}
\text{\bf Riem}[g]=0\,,
\end{equation}
where $\text{\bf Riem}$ denotes the Riemann tensor of $(M,g)$. 
In the second case we regard $g$ as well as the timelike vector 
field $n$ as `dynamical' and add (\ref{eq:RiemZero}) and the 
two equations 
\begin{subequations}
\label{eq:EqForN}
\begin{alignat}{2}
\label{eq:EqForN-1}
& g(n,n)&&\,=\,c^2\,,\\
\label{eq:EqForN-2}
&\nabla n&&\,=\,0\,.
\end{alignat}   
\end{subequations}
In this fashion we arrive at diffeomorphism invariant equations.
But do they really represent the same theory as the one we 
originally started from? For example, are their solution spaces 
`the same'? Naively the answer is clearly `no', simply because the 
reformulated theory has---by construction---a much larger space of 
solutions. For any solution $\Phi$ of the original equations 
$F[\Phi,\Sigma]=0$, where $\Sigma$ is fixed, we now have the whole 
$\diff$--orbit of solutions, 
$\{(\diffmap\cdot\Phi,\diffmap\cdot\Sigma)\mid\diffmap\in\diff\}$
of the new equations, which treat $\Sigma$ as dynamical variable. 
A bijective correspondence can only be established if the 
transformations $\diffmap$ that act non-trivially on $\Sigma$ 
(i.e. $\diffmap\not\in\stab_\diff(\Sigma)$) are declared to be 
\emph{gauge transformations}, so that any two field configurations 
related by such a $\diffmap$ are considered to be physically 
identical. 

If this is done, the simple strategy outlined here suffices to 
(formally) trivialize the 
requirement of diffeomorphism invariance. Hence defining background 
independence as being simple diffeomorphism invariance would also 
render it a trivial requirement. How could we improve its 
definition so as to make it a useful notion? This is precisely what 
Anderson attempted in \cite{Anderson:RelativityPhysics}. He noted 
the following peculiarities of the reformulation just given:
\begin{itemize}
\item[1.]
The new fields $g$ or $(g,n)$ obey an autonomous set of equations 
which does not involve the proper dynamical fields $F$ or $\phi$
respectively. In contrast, the equations for the latter \emph{do} 
involve $g$ or $(g,n)$. Physically speaking, the system whose 
states are parameterized by the new variables acts upon the system 
whose states are parameterized by $F$ or $\phi$, but not vice 
versa. An agent which dynamically acts but is not acted upon 
may well be called `absolute' -- in generalization 
of Newton's absolute space. Such an absolute agent should be 
eliminated.
\item[2.]
The sector of solution space parameterized by $g$ or $(g,n)$
consists of a single diffeomorphism orbit. For example, this 
means that for any two solutions $(\phi,g,n)$ and $(\phi',g',n')$ 
of (\ref{eq:diffusion2}), (\ref{eq:RiemZero}), and (\ref{eq:EqForN}) 
there exists a diffeomorphism $\diffmap$ such that 
$(g',n')=(\diffmap\cdot g\,,\,\diffmap\cdot n)$. 
So `up to diffeomorphisms' there exists only one solution in the 
$(g,n)-$sector. This is far from true for $\phi$: the two solutions $\phi$ 
and $\phi'$ are generally not related by a diffeomorphism. This 
difference just highlights the fact that the added variables really 
did not correspond to new degrees of freedom (they were never supposed 
to) because the added equations were chosen strong enough to maximally 
fix their values (up to diffeomorphisms). 
\end{itemize}

A closer analysis shows that the first criterion is really too much 
dependent on the presentation to be generally useful as a necessary 
condition. Absolute structures will not always reveal their nature 
by obeying autonomous equations. The second criterion is more 
promising and actually entered the literature with some refinements 
as criterion for absolute structures. Before going into this, we will 
discuss some attempts to disable the trivialization strategies just 
outlined.

\subsection{Strategies  against triviality}
\subsubsection{Involving the principle of equivalence}
As diffeomorphism \emph{covariance} is a rather trivial requirement 
to satisfy, we will from now on only be concerned with diffeomorphism 
\emph{invariance}. As we explained, it could be achieved by letting 
the $\Sigma$'s `change sides', i.e. become dynamical structures 
($\gamma$'s and $\Phi$'s), as schematically written down in 
(\ref{eq:trivialize}). We seek sensible criteria that will limit 
the number of such renegades. A physical criterion that suggests 
itself is to allow only those $\Sigma$ to change sides which are 
known to correspond to dynamical variables in a wider context. 
For example, we may allow the spacetime metric $g$ to become 
formally dynamical, since we know that it describes the 
gravitational field, even if in the context at hand the self-dynamics 
of the gravitational field is not relevant and therefore, as a matter 
of approximation, fixed to some value (e.g. the Minkowski metric). 
Doing this would render the Maxwell equations (\ref{eq:Maxwell}) 
(plus the equations for $g$) diffeomorphism invariant. But this 
alone would not work for the diffusion equation, where $n$ would still 
act as a non-dynamical structure. 

Hence we see that the requirement to achieve diffeomorphism 
invariance by at most adjoining $g$ to the dynamical variables is
rather non trivial and connects to Einstein's principle of 
equivalence. Let us quote Wolfgang Pauli in this context
(\cite{Pauli:2000}, p.\,181, his emphasis): 
\begin{quote}
``Einen physikalischen Inhalt bekommt die allgemeine kovariante 
Formulierung der Naturgesetze erst durch das \"Aquivalenzprinzip, 
welches zur Folge hat, da\ss\ die Gravitation durch die $g_{ik}$ 
\emph{allein} beschrieben wird, und das diese nicht unabh\"angig 
von der Materie gegeben, sondern selbst durch die Feldgleichungen 
bestimmt sind. Erst deshalb k\"onnen die $g_{ik}$ als 
\emph{physikalische Zustandsgr\"o\ss en} bezeichnet werden''.\footnote{%
``The generally covariant formulation of the physical laws acquires 
a physical content only through the principle of equivalence, in 
consequence of which gravitation is described \emph{solely} by the 
$g_{ik}$ and these latter are not given independently from matter,
but are themselves determined by field equations. Only for this 
reason can the $g_{ik}$ be described as \emph{physical quantities}''
(\cite{Pauli:ToR-Dover}, p.\,150).}  
(\cite{Pauli:2000}, p.\,181; the emphases are Pauli's)  
\end{quote}

\subsubsection{Absolute structures}
As already remarked, another strategy to render the requirement of 
diffeomorphism invariance non-trivial was suggested by 
Anderson~\cite{Anderson:RelativityPhysics} by means of his notion 
of `absolute structures'. However, most commentators share the 
opinion that Anderson did not succeed to give a proper definition 
of this term. Even worse, some feel that so far nobody has, in fact, 
succeeded in giving a fully satisfying definition. 

To see what is behind this somewhat unhappy state of affairs let us 
start with a tentative definition that suggests itself from the 
discussion given above: 
\begin{definition}[Tentative]
\label{def:AbsStrucTent}
Any field which is either not dynamical, or whose solution space 
consists of a single $\diff$-orbit, is called an 
\textbf{absolute structure}.
\end{definition}    
In general terms, let $\Scal$ denote the space of solutions to a given 
theory. If the theory is $\diff$ invariant $\Scal$ carries an action 
of $\diff$. The fields can be thought of as coordinate functions on 
$\Scal$. An absolute structure is a coordinate which takes the same 
range of values in each $\diff$ orbit and therefore cannot separate 
any two of them. If we regard $\diff$ as a gauge group, i.e. that 
$\diff$--related configurations are physically indistinguishable, 
then absolute structures carry no observable content.  

Following our general strategy we could now attempt to give a 
definition of `background independence': 
\begin{definition}(Tentative)
\label{def:BackIndepTent}
A theory is called \textbf{background independent} iff its equations are 
$\diff$-invariant in the sense of Definition\,\ref{def:diff-invariance}
and its fields do not include absolute structures in the sense of 
Definition\,\ref{def:AbsStrucTent}.
\end{definition}
\noindent
Before discussing these proposal, let us look at some more examples.

\subsection{More examples}
\subsubsection{Scalar gravity a la Einstein-Fokker}
In 1913, just before the advent of General Relativity, Gunnar Nordsr\"om 
invented a formally consistent Poincar\'e--invariant scalar theory of 
gravity, a variant of which we will describe in some detail in the 
second part of this contribution.\footnote{In fact, there are two 
related but inequivalent scalar theories by Nordsr\"om; see e.g. 
\cite{Norton:1992}. The one presented in part\,2 is essentially 
equivalent to a theory sketched by Otto Bergmann in 1956 
\cite{Bergmann:1956}, which Harvey \cite{Harvey:1955} classified 
as a modification of Nordstr\"oms first theory.}   
Its essence is the field equation (\ref{eq:EqScalGrav1}) and the 
equation of motion (\ref{eq:ParticleMotion1}) for a test particle. 
Shortly after its publication it was pointed out by Einstein and Fokker 
that Nordstr\"om's (second) theory can be presented in a `covariant' way. 
Explicitly they said: 
\begin{quote}
``Im folgenden soll dargetan werden, da\ss\ man zu einer in formaler 
Hinsicht vollkommen geschlossenen und befriedigenden Darstellung der 
Theorie [Nordstr\"oms] gelangen kann, wenn man, wie dies bei der 
Einstein-Grossmannschen Theorie bereits geschehen ist, das 
invarianten-theoretische Hilfsmittel benutzt, welches uns in dem 
absoluten Differentialkalk\"ul gegeben ist''.\footnote{%
``In the following we wish to show that one can arrive at a formally 
complete and satisfying presentation of the theory [Nordstr\"om's] 
if one uses the methods from the theory of invariants given by the 
absolute differential calculus, as it was already done in the 
Einstein-Grossman theory''.} (\cite{CPAE}, Vol.\,4, Doc.\,28, p.\,321)
\end{quote}
The essential observation is this: consider conformally flat 
metrics:
\begin{equation}
\label{eq:Nordsrom3}
g_{\mu\nu}=\phi^2\,\eta_{\mu\nu}\,,
\end{equation}
then the field equation is equivalent to 
\begin{subequations}
\label{eq:EinsteinFokker}
\begin{equation}
\label{eq:EinsteinFokker-1}
R[g]  \,=\, 24\pi G\,g^{\mu\nu}T_{\mu\nu}\,,
\end{equation}
where $R[g]$ is the Ricci scalar for the metric $g$,
whereas the equation of motion for the particle becomes the 
geodesic equation with respect to $g$:
\begin{equation}
\label{eq:EinsteinFokker-2}
\ddot x^\mu+\Gamma^\mu_{\alpha\beta}\dot x^\alpha \dot x^\beta=0\,.
\end{equation}
\end{subequations}
Now, the system (\ref{eq:EinsteinFokker}), considered as equations 
for the metric $g$ and the trajectory $x$, is clearly 
$\diff$-invariant. But Nordstr\"oms theory is equivalent to 
(\ref{eq:EinsteinFokker}) \emph{plus} (\ref{eq:Nordsrom3}). Here 
$\eta$ is a non-dynamical field so that 
(\ref{eq:EinsteinFokker},\ref{eq:Nordsrom3}) is only $\diff$-covariant. 
According to the general scheme outlined above this could be remedied 
by letting the metric $\eta$ be a new dynamical variable whose 
equation of motion just asserts its flatness: 
\begin{equation}
\label{eq:Nordstrom6}
\text{\bf Riem}[\eta]=0\,.  
\end{equation}
But then $\eta$ qualifies as an absolute structure 
according to Definition\,\ref{def:AbsStrucTent} and the theory 
(\ref{eq:EinsteinFokker},\ref{eq:Nordsrom3},\ref{eq:Nordstrom6}) 
is not background independent. The subgroup $G\subset\diff$ that 
stabilizes $\eta$ is---by definition--- the inhomogeneous Lorentz
group, which had already been the invariance group of Nordstr\"oms 
theory. So no additional invariance has, in fact, been gained in 
the transition from Nordstr\"om's to the Einstein-Fokker formulation. 

Sometimes the absolute structures are not so easy to find because 
the theory is formulated in such a way that they are not yet isolated 
as separate field. For example, in the case at hand, (\ref{eq:Nordsrom3}) 
and (\ref{eq:Nordstrom6}) together are clearly equivalent to the single 
condition that $g$ be conformally flat, which in turn is 
equivalent to the vanishing of the conformal curvature tensor for  
$g$ (Weyl tensor): 
\begin{equation}
\label{eq:Nordstrom7}
\text{\bf Weyl}[g]=0\,.  
\end{equation}
The field $\eta_{\mu\nu}$ has now disappeared from the description and 
the theory does not explicitly display any absolute structure anymore. 
But, of course, it is still there; it is now part of the field 
$g$. To bring it back to light, make a field redefinition 
$g_{\mu\nu}\mapsto (\phi,h_{\mu\nu})$ which isolates the part 
determined by (\ref{eq:Nordstrom7}); for example   
\begin{alignat}{2}
\label{eq:Nordstrom8}
& \phi      &&:=\left[-\det\{g_{\mu\nu}\}\right]^{\frac{1}{8}}\,,\\
\label{eq:Nordstrom9}
& h_{\mu\nu}&&:=g_{\mu\nu}\,
\left[-\det\{g_{\mu\nu}\}\right]^{-\frac{1}{4}}\,.
\end{alignat}
Then any two solutions for the full set of equations are such 
that their component fields $h_{\mu\nu}$ and $h'_{\mu\nu}$
are related by a diffeomorphism. Hence $h_{\mu\nu}$ 
is an absolute structure. 

Clearly there is a rather non-trivial mathematical theory behind 
the last statement of diffeomorphism equivalence of $h_{\mu\nu}$.
We could not have made that statement had we not already been in 
possession of the full solution theory for (\ref{eq:Nordstrom7}) 
which, after all, is a complicated set of non-linear partial 
differential equations of second order.

\subsubsection{A massless scalar field from an action principle}
Usually we require the equations of motion to be the 
Euler-Lagrange equations for some associated action principle. 
Would the somewhat bold strategy to render non-dynamical structures 
dynamical by adding \emph{by hand} `equations of motion' which fix 
them to their previous values also work if these added equations 
were required to be the Euler-Lagrange equations for some common 
action principle? The answer is by no means obvious, as the following 
simple example taken from \cite{Sorkin:2002} illustrates: 

Consider a real massless\footnote{This is just assumed for simplicity. 
The arguments works the same way if a mass term were included.} 
scalar field in Minkowski space:
\begin{equation}
\label{eq:Sorkin-1}
\square\phi:=\eta^{\mu\nu}\nabla_\mu\nabla_\nu\phi=0\,.
\end{equation}
According to standard strategy the non-dynamical Minkowski metric 
$\eta$ is eliminated by introducing the dynamical variable $g$,
replacing $\eta$ in (\ref{eq:Sorkin-1}) by $g$, and adding the 
flatness condition 
\begin{equation}
\label{eq:Sorkin-2}
\text{Riem}[g]=0
\end{equation}
as new equation of motion. Is there an action principle whose 
Euler-Lagrange equations are (equivalent to) these equations?
This seems impossible without introducing yet another field $\lambda$
(a Lagrange multiplier) whose variation just yields 
(\ref{eq:Sorkin-2}). The action would then be
\begin{equation}
\label{eq:Sorkin-3}
S = \tfrac{1}{2}\int dV\ g^{\mu\nu}\nabla_\mu\phi\,\nabla_\nu\phi
+ \tfrac{1}{4}\int dV\ \lambda^{\alpha\beta\mu\nu}R_{\alpha\beta\mu\nu}\,,
\end{equation}
where the symmetries of the tensor field $\lambda$ are that of 
the Riemann tensor: 
\begin{equation}
\label{eq:Sorkin-4}
\lambda^{\alpha\beta\mu\nu}=
\lambda^{[\alpha\beta][\mu\nu]}=
\lambda^{\mu\nu\alpha\beta}\,.
\end{equation}
Variation with respect to $\phi$ and $\lambda$ yield (\ref{eq:Sorkin-1})
and (\ref{eq:Sorkin-2}) respectively, and variation with respect to 
$g$ gives 
\begin{equation}
\label{eq:Sorkin-5}
\nabla_\mu\nabla_\nu\lambda^{\alpha\mu\beta\nu}=T^{\alpha\beta}\,,  
\end{equation}
where $T^{\alpha\beta}$ is the energy-momentum tensor for $\phi$.
These equations do not give a background independent theory for 
the fields $(\phi,g,\lambda)$ since $g$ is an absolute structure. 
The solution manifold of the $\phi$ field is, in fact, the same 
as before. For this it is important to note that there is an 
integrability condition resulting from (\ref{eq:Sorkin-5},\ref{eq:Sorkin-2}),
namely $\nabla_\alpha T^{\alpha\beta}=0$, which is however already 
implied by (\ref{eq:Sorkin-1}). Hence no extra constraints on $\phi$ 
result from (\ref{eq:Sorkin-5}). 

However, the $\lambda$ field seems to actually add more 
dimensions to the solution manifold and hence to the observable 
content of the theory. Indeed, using the Poincar\'e Lemma in 
flat space one shows that any divergenceless symmetric 2-tensor 
$T^{\mu\nu}$ can always be written as in (\ref{eq:Sorkin-5}), 
where $\lambda$ has the symmetries (\ref{eq:Sorkin-4}). 
But this does not fix $\lambda^{\mu\alpha\nu\beta}$, so that 
the set of $\diff$--equivalence classes of stationary points of 
(\ref{eq:Sorkin-3}) is strictly `larger' than the set of solutions 
of (\ref{eq:Sorkin-1}). In other words, the ($\diff$ reduced) phase 
space for the theory described by (\ref{eq:Sorkin-3}) is `larger' 
then that for (\ref{eq:Sorkin-1}).\footnote{I am not aware of a 
reference where a Hamiltonian reduction of (\ref{eq:Sorkin-3}) is 
carried out.} A a result we conclude that the 
reformulation given here does \emph{not} achieve an equivalent 
$\diff$--invariant reformulation of (\ref{eq:Sorkin-1}) in terms 
of an action principle.   

\subsection{Problems with absolute structures}
\label{sec:ProblemsAbsStr}
A first thing to realize form the examples above is that the notion of 
absolute structure should be slightly refined. More precisely, it should
be made local in order to capture the idea that an absolute element 
in the theory does not represent local degrees of freedom. Rather than 
saying that a field corresponds to an absolute structure if its solution 
space consists of a single $\diff$--orbit, we would like to make the 
latter condition local:
\begin{definition}
\label{def:LocallyDiffEquiv}
Two fields $T_1$ and $T_2$ are said to be 
\textbf{locally diffeomorphism equivalent} iff for any point 
$p\in M$ there exits a neighbourhoods $U$ of $p$ and a 
diffeomorphism $\phi_U:U\rightarrow U$ such that 
$\phi_U\cdot(T_1\big\vert_{U})=T_2\big\vert_{U}$.
\end{definition}
Note that local diffeomorphism equivalence defines an equivalence 
relation on the set of fields. Accordingly, following a suggestion of 
Friedman~\cite{Friedman:1973}, we should replace the tentative 
Definition\,\ref{def:AbsStrucTent} by the following

\begin{definition}
\label{def:AbsStruc}
Any field which is either not dynamical or whose solutions are all 
locally diffeomorphism equivalent is called an 
\textbf{absolute structure}.
\end{definition}    
In fact, this is what we implicitly used in the discussions above
where we slightly oversimplified matters. For example, any two 
flat metrics $g_1,g_2$ (i.e. which satisfy $\text{\bf Riem}[g_{1,2}]=0$) 
are generally only \emph{locally} diffeomorphism equivalent. Likewise, 
a conformally flat metric $g$ (i.e. which satisfy $\text{\bf Weyl[g]=0}$) 
is \emph{locally} diffeomorphism equivalent to $f^2 \eta$, where $f$ is 
non-vanishing function and $\eta$ is a fixed flat metric. 

Having corrected this we should also adapt the tentative 
Definition\,\ref{def:BackIndepTent}:  
\begin{definition}
\label{def:BackIndep}
A theory is called \textbf{background independent} iff its equations are 
$\diff$-invariant in the sense of Definition\,\ref{def:diff-invariance}
and its fields do not include absolute structures in the sense of 
Definition\,\ref{def:AbsStruc}.
\end{definition}

So far so good. Is this, then, the final answer? Unfortunately not!
The standard argument against \emph{this} notion of absolute structure 
is that it may render structures absolute that one would normally 
call dynamical. The canonical example, usually attributed to 
Robert~Geroch~\cite{Jones:1981}, makes use of the well known fact 
in differential geometry that nowhere vanishing vector fields are 
always locally diffeomorphism equivalent (see e.g. Theorem 2.1.9 
in \cite{AbrahamMarsden:Mechanics}). Hence any diffeomorphism invariant 
theory containing vector fields among their fundamental field variables 
cannot be background independent. For example, consider the coupled 
Einstein-Euler equations for a perfect fluid  of density $\rho$ and 
four-velocity $u$ in spacetime with metric $g$. This system of 
equations is $\diff$-invariant. By definition of a velocity field 
we have $g(u,u)=c^2$. This means that $u$ cannot have zeros, even if 
for physical reasons we would usually assume the fluid to be not 
everywhere in spacetime, i.e. the support of $\rho$ is a proper 
subset of spacetime.\footnote{It seems a little strange to be forced 
to consider velocity fields $u$ in regions where $\rho=0$, i.e. where 
there is no fluid matter. Velocity of what? one might ask. 
In concrete applications this means that we have to extend $u$ 
beyond the support of $\rho$ and that the physical prediction is 
independent of that extension.} Then the four velocity of the fluid 
is an absolute structure, contrary to our physical intention. 

I know of two suggestions how to avoid this conclusion in the 
present example. One is to use the 1-form $u_\mu\,dx^\mu$ rather 
than the vector field $u^\mu\partial_\mu$ as fundamental dynamical 
variable for the fluid. The point being that one-form fields are 
\emph{not} locally diffeomorphism equivalent. For example, a closed 
(exact) one-form field will always be mapped into a closed (exact) 
one-form field, and hence cannot be locally diffeomorphism equivalent 
to a non-closed field. Another suggestion, in fact the only one that 
I have seen in the literature (\cite{Friedman:1983} p.\,59 footnote\,9
and \cite{Straumann:GR}, p.\,99, footnote\,8) is to take the 
energy-momentum density $\Pi$ rather than $u$ as fundamental variable. 
To be sure, on the support of $\Pi$ we can think of it as equal 
to $\rho u$, but on the complement of its support there is no need 
to define a $u$.  This avoids the unwanted conclusion whenever $\Pi$ 
indeed has zeros; otherwise the argument given above for $u$ just
applies to $\Pi$. 

An even simpler argument, which I have not seen in the physics 
literature, even applies to pure gravity. It rests on the following 
theorem from differential geometry, an elegant proof of which was given 
by Moser~\cite{Moser:1965}: given two compact oriented $n$-dimensional 
manifolds $V_1$ and $V_2$ with $n$-forms $\mu_1$ and $\mu_2$ respectively. 
There exists an orientation preserving diffeomorphism 
$\phi:V_1\rightarrow V_2$ such that $\phi^*\mu_2=\mu_1$ iff the 
$\mu_1$-volume of $V_1$ equals the $\mu_2$-volume of $V_2$, i.e. iff
\begin{equation}
\label{eq:ThmMoser1}
\int_{V_1}\mu_1=\int_{V_2}\mu_2\,.
\end{equation}
If we take $V_1=V_2$ to be the closure of an open neighbourhood $U$ 
in the spacetime manifold $M$, this theorem implies that the metric 
volume forms, written in coordinates as  
\begin{equation}
\label{eq:ThmMoser2}
\mu=\sqrt{\big\vert\det[g(\partial_\mu,\partial_\nu)]\big\vert}\, 
dx^1\wedge\cdots\wedge dx^n\,,
\end{equation}
are locally diffeomorphism equivalent iff they assign the same volume 
to $U$. Hence it follows that the metric volume elements modulo 
constant factors are absolute elements in pure gravity. Note that
this implies that for any metric $g$ any any point $p\in M$ there 
is always a local coordinate system $\{x^\mu\}$ in an open 
neighbourhood $U$ of $p$ such that 
$\sqrt{\vert\det[g(\partial_\mu,\partial_\nu)]\vert}=1$. 

\newpage
\section{A historical note on scalar gravity}
In his contribution (``Physikalischer Teil'') to the `Entwurf Paper' 
(\cite{CPAE}, Vol.\,4, Doc.\,13), that Einstein wrote with his lifelong 
friend Marcel Grossmann\footnote{Marcel Grossmann wrote the 
``Mathematischer Teil''.}, Einstein finished with \S\,7 whose title 
asks: ``Can the gravitational field be reduced to a scalar\,?''
(``Kann das Gravitationsfeld auf einen Skalar zur\"uckgef\"uhrt werden\,?'').
There he presented a Gedankenexperiment-based argument which apparently 
shows that any Poincar\'e-invariant\footnote{By `Poincar\'e group' we 
shall understand the inhomogeneous $SL(2,\complex)$, i.e. the semi-direct 
product $\reals^4\rtimes SL(2,\complex)$, defined by the multiplication 
law $(a,A)(b,B)=(a+\pi(A)b,AB)$, where 
$\pi:SL(2,\complex)\rightarrow SO(1,3)_0$ (the identity component of 
$SO(1,3)$) is the 2-1 covering homomorphism. The phrase 
`Poincar\'e-invariance' is always taken to mean that the equations of 
motion admit the Poincar\'e group as symmetry group, i.e. it transforms 
solutions to solutions of the \emph{very same} equation.} scalar theory 
of gravity, in which the scalar gravitational field couples exclusively 
to the trace of the energy-momentum tensor, necessarily violates 
energy conservation and is hence physically inconsistent. This he 
presented as plausibility argument why gravity has to be described by
a more complex quantity, like the $g_{\mu\nu}$ of the `Entwurf Paper', 
where he and Grossmann considers `generally covariant' equations for the 
first time. After having presented his argument, he ends \S\,7 (and his 
contribution) with the following sentences, showing that his 
conviction actually derived on some form of the PGC: 
\begin{quote}
``Ich mu\ss\ freilich zugeben, da\ss\ f\"ur mich das wirksamste Argument 
dar\"ur, da\ss\ eine derartige Theorie [eine skalare Gravitationstheorie]
zu verwerfen sei, auf der \"Uberzeugung beruht, da\ss\ die Relativit\"at 
nicht nur orthogonalen linearen Substitutionen gegen\"uber besteht, 
sondern einer viel weitere Substitutionsgruppe gegen\"uber. Aber wir 
sind schon desshalb nicht berechtigt, dieses Argument geltend zu machen, 
weil wir nicht imstande waren, die (allgemeinste) Substitutionsgruppe 
ausfindig zu machen, welche zu unseren Gravitationsgleichungen 
geh\"ort''.\footnote{%
To be sure, I have to admit that in my opinion the most effective 
argument for why such a theory [a scalar theory of gravity] has to 
be abandoned rests on the conviction that relativity holds with 
respect to a much wider group of substitutions than just the 
linear-orthogonal ones.  However, we are not justified to push 
this argument since we were not able to determine the (most general) 
group of substitutions which belongs to our gravitational equations.}
(\cite{CPAE}, Vol.\,4, Doc.\,13, p.\,323)
\end{quote}

Einstein belief, that scalar theories of gravity are ruled out, 
placed him---in this respect---in opposition to most of his 
contemporary physicist who took part in the search for a (special-) 
relativistic theory of gravity (Nordstr\"om,  Abraham, Mie, von~Laue ..).  
Some of them were not convinced, it seems, by Einstein's inconsistency 
argument. For example, even after General Relativity was completed,  
Max von Laue wrote a comprehensive review paper on Nordstr\"oms theory, 
thereby at least implicitly claiming inner consistency \cite{Laue:1917}. 

On the other hand, modern commentators seem to fully accept Einstein's 
claim and view it as important step in the development of General 
Relativity \cite{Norton:1992}\cite{Norton:1993a} and possibly also 
as an important step towards the requirement of general covariance.     
From a modern field theoretic viewpoint, however, the claim of 
violation of energy conservation of a Poincar\'e-invariant theory 
sounds even paradoxical, since Noether's theorem guarantees the 
existence of a conserved quantity associated to the symmetry of 
time-translations. This quantity is usually identified with energy 
(or even taken as definition of energy). Hence Einstein's argument 
cannot be entirely obvious. It even becomes intrinsically incorrect 
if placed within a straightforward scalar theory of gravity, as 
will be shown below. 

\subsection{Einstein's argument}
Einstein first pointed out that the source for the gravitational
field must be a scalar built from the matter quantities alone, and 
that the only such scalar is the trace $T^{\mu}_{\mu}$ of the 
energy-momentum tensor (as pointed out to Einstein by von~Laue, 
as Einstein acknowledges, calling  $T^{\mu}_{\mu}$ the ``Laue Scalar''). 
Moreover, for closed stationary systems the so-called Laue-Theorem 
states that the integral over space of $T^{\mu\nu}$ must vanish, 
except for $\mu=0=\nu$; hence the space integral of $T^{\mu}_{\mu}$ 
equals that of $T^{00}$, which means that the total (active and passive) 
gravitational mass of a closed static system equals its inertial mass. 
However, if the system is not closed, the weight depends on the 
stresses (the spatial components $T^{ij}$), which Einstein deems 
unacceptable.
\begin{wrapfigure}{r}{0.3\linewidth}
\vspace{-0.3cm}
\centering\epsfig{figure=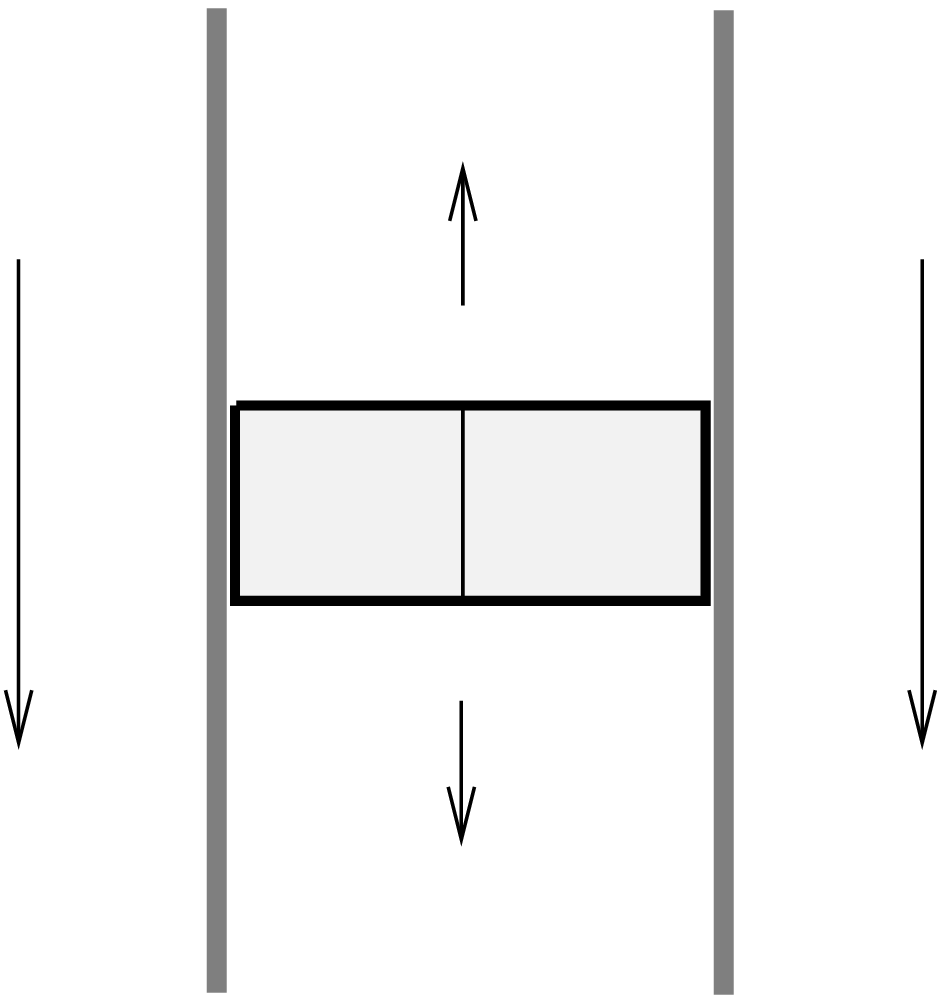,width=\linewidth}
\put(-5,17){\small $\vec g$}
\put(-55,55){\tiny -- strut}
\put(-70,54){\small\textit{B}}
\put(-45,126){\small shaft}
\put(-26.3,117){$\vert$}
\end{wrapfigure}
His argument proper is then as follows: consider a box $B$ filled 
with electromagnetic radiation of total energy $E$. We idealize the walls 
of the box to be inwardly perfectly mirrored and of infinite stiffness, 
i.e. they can support normal stresses (pressure) without any deformation. 
The box has an additional vertical strut in the middle connecting 
top and bottom walls, which supports all the vertical material 
stresses that counterbalance the radiation pressure, so that the 
side walls merely sustain normal and no tangential stresses. 
The box can slide without friction along a vertical shaft, $S$, whose 
cross section corresponds exactly to that of the box. The walls of 
the shaft are likewise idealized to be inwardly perfectly mirrored 
and of infinite stiffness. The whole system of shaft and box is 
finally placed in a homogeneous static gravitational field, $\vec g$,
which points vertically downward. Now we perform the following 
process. We start with the box being 
placed in the shaft in the upper position. Then we slide it 
down to the lower position; see Fig.\,1. There we remove the 
side walls of the box---without any radiation leaking out---such 
that the sideways pressures are now provided by the shaft walls. 
The strut in the middle is left in position to further take all the 
vertical stresses, as before. Then the box together with the 
detached side walls are pulled up to their original positions. 
Finally the system is reassembled so that it assumes its initial state. 
Einstein's claim is now that in a very general class of imaginable 
scalar theories the process of pulling up the parts needs less work 
than what is gained in energy in letting the box (with side walls 
attached ) down. Hence he concluded that such theories necessarily 
violate energy conservation.
\noindent
\begin{figure}
\hspace{1.0cm}
\begin{minipage}[b]{0.25\linewidth}
\centering\epsfig{figure=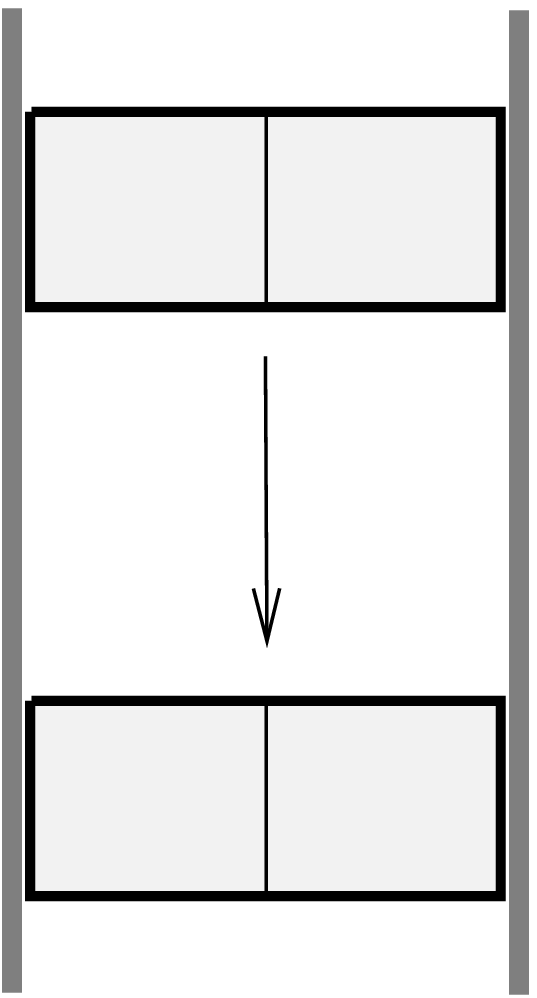,width=\linewidth}
\caption{\small Lowering the box in the gravitational field 
         with side walls attached.}
\end{minipage}\hfill
\begin{minipage}[b]{0.28\linewidth}
\centering\epsfig{figure=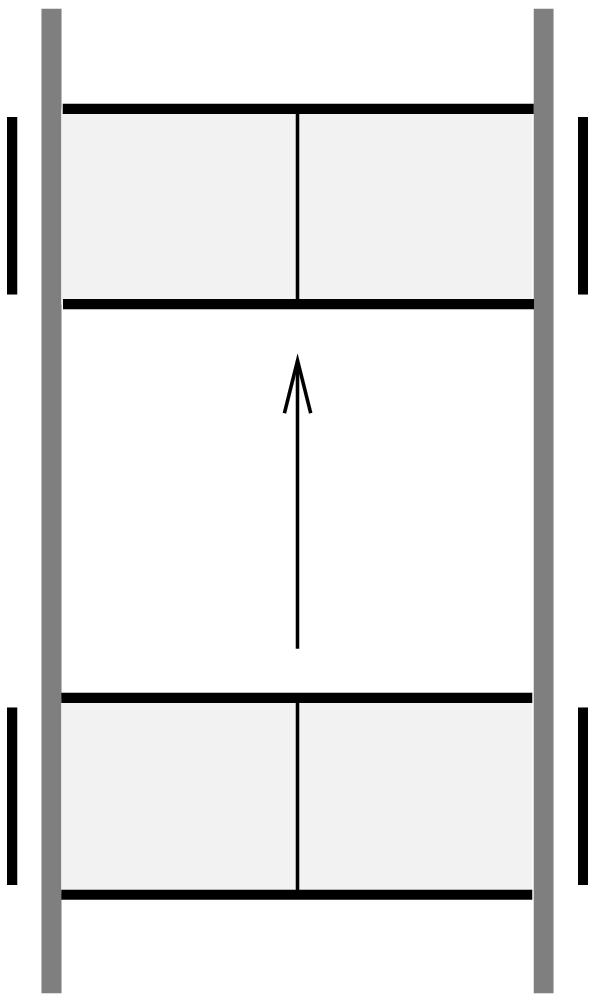,width=\linewidth}
\caption{\small Raising the box in the gravitational field 
         with side walls taken off.}
\end{minipage}
\hspace{1.0cm}
\end{figure}

Indeed,  radiation plus box is a closed static system. Hence the weight 
of the total system is proportional to its total energy $E$, which we 
may pretend to be given by the radiation energy alone, since the 
contributions from the rest masses of the walls will cancel in the 
final energy balance, so that we may formally set them to zero at 
this point. Lowering this box by an amount $h$ in a static homogeneous 
gravitational field of strength $g$ results in an energy gain of 
$\Delta E=hgE/c^2$. So despite the fact that radiation has a traceless 
energy-momentum tensor, \emph{trapped} radiation has a weight given by 
$E/c^2$. This is due to the radiation pressure which puts the walls 
of the trapping box under tension. Tension makes an independent 
contribution to weight, independent of the material that supports 
it.  For each parallel pair of side-walls the tension is just the 
radiation pressure, which is one third of the energy density. 
So each pair of side-walls contribute $E/3c^2$ to the (passive) 
gravitational mass (over and above their rest mass, which we set 
to zero) in the lowering process when stressed, and zero in the 
raising process when unstressed. Hence, Einstein concluded, there 
is a net gain in energy of $2E/3c^3$ (there are two pairs of side 
walls). 

But it seems that Einstein neglects the fact that, in contrast to 
the lowering process, during the lifting process the state of the 
shaft $S$ \emph{is} changed. Moreover, the associated contribution 
to the energy balance just renders Einstein's argument inconclusive. 
Indeed, when the side walls are first removed in the lower position, 
the walls of the shaft necessarily come under stress because they 
now need to provide the horizontal balancing pressures. In the raising 
process that stress distribution of the shaft is translated upwards. 
But that \emph{does} cost energy, even though it is not associated 
with any proper transport of the material the shaft is made from. 
As already pointed out, stresses make their own contribution to weight, 
independent of the nature of the material that supports them. 
In particular, a redistribution of stresses in a material immersed 
in a gravitational field will generally makes a non-vanishing 
contribution to the energy balance, even if the material does not 
move. This is explicitly seen in the model theory discussed next.

\subsection{A formally consistent  model-theory for scalar gravity}  
We wish to construct a Poincar\'e-invariant theory of a scalar 
gravitational field, $\Phi$, coupled to matter. We will use 
Lagrangian methods. Regarding the Minkowski metric we use the 
`mostly minus' convention, that is, 
$\eta_{\mu\nu}=\mbox{diag}(1,-1,-1,-1)$. 

We start from the obvious generalization of Poisson's equation, 
$\Delta\Phi=4\pi G\rho$, with the `Laue-scalar' as source:
\begin{equation}
\label{eq:EqScalGrav1}
\square\Phi=-\kappa T\,,\quad\text{where}\quad\kappa:=4\pi G/c^2\,.
\end{equation}
Here $\square=\eta^{\mu\nu}\partial_{\mu}\partial_{\nu}$ and 
$T=\eta^{\mu\nu}T_{\mu\nu}$. $T_{\mu\nu}$ is the stress-energy 
tensor of the matter (sign-normalization: $T_{00}=T_0^0=+$energy-density). 
We seek an action which makes (\ref{eq:EqScalGrav1}) its 
Euler-Lagrange equation. It's easy to guess\footnote{Note that 
$\Phi$ has the physical dimension of a squared velocity, $\kappa$ 
that of length-over-mass. The prefactor $1/\kappa c^3$ gives 
the right hand side of (\ref{eq:ActionFieldInt}) the physical 
dimension of an action. The overall signs are chosen according to 
the general scheme for Lagrangians: kinetic minus potential energy.}:
\begin{equation}
\label{eq:ActionFieldInt}
S_{\rm field}+S_{\rm int}= \frac{1}{\kappa c^3}\int d^4x\left(
\tfrac{1}{2}\partial_\mu\Phi\partial^\mu\Phi-\kappa\Phi T\right)\,.
\end{equation}
where $S_{\rm field}$, given by the first term,  is the 
action for the gravitational field and $S_{\rm int}$, 
given by the second term, accounts for the interaction with matter.
 
To this we have to add the action $S_{\rm matter}$ for the matter,
which we only specify insofar as we we assume that the matter consists
of a point particle of rest-mass $m_0$ and a `rest' that needs not be
specified further for our purposes here.  
Hence $S_{\rm matter}=S_{\rm particle}+ S_{\rm rom}$ 
(rom = rest of matter) where
\begin{equation}
\label{eq:ActionPP}
S_{\rm particle}=
-m_0c^2\int d\tau\,.
\end{equation}
The quantity $d\tau=\tfrac{1}{c}\sqrt{\eta_{\mu\nu}dz^\mu dz^\nu}$ is the
proper time along the worldline of the particle. The energy-momentum
tensor of the particle is given by
\begin{equation}
\label{eq:T-PointParticle}
T^{\mu\nu}(x)=m_0c\,\int 
{\dot z}^\mu(\tau){\dot z}^\nu(\tau)\ \delta^{(4)}(x-z(\tau))\ d\tau\,,
\end{equation}
so that the particle's contribution to the interaction term in
(\ref{eq:ActionFieldInt}) is
\begin{equation}
\label{eq:InteractionPP}
S_{\text{int-particle}}=-m_0\int\Phi(z(\tau))\ d\tau\,.
\end{equation}
Hence the total action can be written in the following form:
\begin{equation}
\label{eq:ScalGrav4}
\begin{split}
S_{\rm tot}=
& - m_0c^2\int\bigl(1+\Phi(z(\tau))/c^2\bigr)\ d\tau\\
& +\frac{1}{\kappa c^3}\int d^4x\ \bigl(\tfrac{1}{2}\partial_\mu\Phi\partial^\mu\Phi
  -\kappa\Phi T_{\text{rom}}\bigr)\\
& +S_{\text{rom}}\,.
\end{split}
\end{equation}

By construction the field equations that follow from this action are 
given by (\ref{eq:EqScalGrav1}), where the energy momentum-tensor refers 
to the matter without the test particle, if we treat the latter as 
\emph{test} particle. The equations of motion for the test particle 
are then given by  
\begin{subequations}
\label{eq:ParticleMotion}
\begin{alignat}{3}
\label{eq:ParticleMotion1}
& &&\ddot z^\mu&&\,=\,P^{\mu\nu}\partial_\nu\phi\,,\\
\label{eq:ParticleMotion2}
&\text{where}\qquad
&&P^{\mu\nu}&&\,=\,\eta^{\mu\nu}-{\dot z}^\mu{\dot z}^\nu/c^2\\
\label{eq:ParticleMotion3}
&\text{and}\qquad
&&\phi&&\,=\,c^2\ln(1+\Phi/c^2)\,.
\end{alignat}
\end{subequations}
Two things are worth remarking at this point: 
\begin{itemize}
\item
The term $P^{\mu\nu}$ is a projector perpendicular to the timelike 
direction given by $\dot z$. It is necessary in order to avoid 
overdetermination. Due to $\dot z^\mu \dot z_\mu=c^2$ there can only be 
three independent equations of motion. Indeed, an equation like 
$\ddot z^\mu=\partial^\mu\phi$ immediately leads to the 
integrability condition $\dot z^\mu\partial_\mu\phi=0$, which 
renders this equations useless since it says that $\phi$ may not 
change along the worldline of the particle.
\item
Whereas $\Phi$ plays the analog of the Newtonian potential in the 
SR-adapted field equation (\ref{eq:EqScalGrav1}), it is $\phi$
rather than $\Phi$ that plays the analog of the Newtonian equation
potential in the equation of motion for a test particle. The relation 
between the two potential is given by (\ref{eq:ParticleMotion3}). 
We were not free to just impose an equation of motion for the 
test particle, in which $\phi$ in (\ref{eq:ParticleMotion1}) is 
replaced by $\Phi$. Rather, (\ref{eq:ParticleMotion1}) is an 
unambiguous consequence of the consistency requirement, according to 
which all forms of matter couple to gravity in the \emph{same}
fashion, namely via the $\Phi T$ -- term in the interaction 
Lagrangian. From (\ref{eq:T-PointParticle}) via 
(\ref{eq:InteractionPP}) this directly leads to 
(\ref{eq:ParticleMotion}).     
\end{itemize}   

Suppose there exists some inertial coordinate system $x^\mu$ 
with respect to which $\Phi$ (and hence $\phi$) is static, 
i.e. $\partial_0\Phi=0$, then in these coordinates 
(\ref{eq:ParticleMotion1}) is equivalent to the following 
3-vector equation ($t=x^0$)
\begin{equation}
\label{eq:ParticleMotion4}
\tfrac{d^2}{dt^2}\vec z(t)=
-\left(1-\vert\tfrac{d}{dt}\vec z(t)\vert^2/c^2\right)\ 
\vec\nabla\varphi(\vec z(t))\,.
\end{equation}
From Einstein's own recollections we know that he also arrived at 
an equation like (\ref{eq:ParticleMotion4}) in an early attempt 
to generalize Newton's scalar theory of gravity, but that he dismissed 
it for not satisfying some variant of the universality of free fall, 
according to which the vertical acceleration of a body should be 
independent of the horizontal velocity of its center of mass. 
In his own words: 
\begin{quote}
``Dieser Satz, der 
auch als Satz \"uber die Gleichheit der tr\"agen und schweren Masse 
formuliert werden kann, leuchtete mir nun in seiner tiefen Bedeutung
ein. Ich wunderte mich im h\"ochsten Grade \"uber sein Bestehen und 
vermutete, dass in ihm der Schl\"ussel f\"ur ein tieferes Verst\"andnis 
der Tr\"agheit und Gravitation liegen m\"usse. An seiner strengen 
G\"ultigkeit habe ich auch ohne Kenntnis des Resultates der sch\"onen 
Versuche von E\"otv\"os, die mir -- wenn ich mich richtig erinnere 
-- erst sp\"ater bekannt wurden, nicht ernsthaft 
gezweifelt.''\footnote{%
``These investigations, however, led to a result which raised my
strong suspicion. According to classical mechanics, the vertical
acceleration of a body in the vertical gravitational field is
independent of the horizontal component of its velocity.
Hence in such a gravitational field the vertical acceleration
of a mechanical system or of its center of gravity comes out
independently of its internal kinetic energy. But in the theory
I advanced, the acceleration of a falling body was not
independent of its horizontal velocity or the internal energy
of the system. This did not fit with the old experimental
fact that all bodies have the same acceleration in a
gravitational field.''} (\cite{Einstein:MeinWeltbild}, pp.\,135--136)
\end{quote}
Concerning this statement, at least three things seem 
truly remarkable: 
\begin{itemize}
\item
That Einstein would dismiss the \emph{quadratic} dependence 
of the vertical acceleration on $v/c$, as predicted by 
(\ref{eq:ParticleMotion4}), as ``not in accord with the `old 
experience' (sic!) of the universality of free fall''.
\item
The dependence of the vertical acceleration on the horizontal
center-of-mass velocity is clearly expressed by 
(\ref{eq:ParticleMotion4}). However, Einstein's additional claim 
that there is also a similar dependence on the internal energy 
does not survive closer scrutiny. One might think at first that 
(\ref{eq:ParticleMotion4}) also predicts that, for example, the 
gravitational acceleration of a box filled with a gas decreases 
with temperature, due to the increasing velocities of the gas 
molecules. But this arguments neglects the walls of the box 
which gain in stress due to the rising gas pressure. According 
to (\ref{eq:EqScalGrav1}) more stress means less weight. In fact, 
a general argument due to Laue (1911) shows that these effects
precisely cancel (see e.g. \cite{Norton:1993a} for a lucid
discussion).
\item
Einstein's requirement that the vertical acceleration should be 
independent of the horizontal velocity is (for good reasons) not 
at all implied by the modern formulation of the (weak) equivalence 
principle, according to which the worldline of a freely falling 
test-body (without higher mass-multipole-moments and without charge 
and spin) is determined by its initial spacetime point and four 
velocity, i.e. independent of the further constitution of the test 
body. In contrast, Einstein's requirement relates two motions with 
\emph{different} initial velocities. In fact, it is badly in need of a 
proper interpretation to even make physical sense. Are we to require 
that two bodies dropped from some altitude, one with the other without 
horizontal initial velocity, reach the ground simultaneously?
What what does `simultaneously' refer to? Simultaneously in the 
initial rest frame of one of the two bodies? Or at the same lapse of 
eigentimes of the two bodies? 
\end{itemize}

\noindent
In passing we remark that (\ref{eq:ParticleMotion4}) gives rise to 
a periastron precession of $-1/6$ times the value obtained from 
GR.

\subsection{Energy conservation}
Corresponding to Poincar\'e-invariance there are 10 conserved currents. 
In particular, the total energy $E$ relative to an inertial system
is conserved. For a particle coupled to gravity it is easily 
calculated and consists of three contributions corresponding to 
the gravitational field, the particle, and the interaction-energy  
of particle and field:
\begin{subequations}
\label{eq:Energies}
\begin{alignat}{2}
\label{eq:EnergiesGrav}
& E_{\sss\rm gravity}
&&\,=\,\frac{1}{2\kappa c^2}\int d^3x\ 
  \bigl((\partial_{ct}\Phi)^2+(\vec\nabla\Phi)^2\bigr)\,,\\
\label{eq:EnergiesPart}
& E_{\sss\rm particle} 
&&\,=\, m_0c^2\,\gamma(v)\,,\\
\label{eq:EnergiesInt}
& E_{\sss\rm interaction} 
&&\,=\, m_0\,\gamma(v)\,\Phi\bigl(\vec z(t),t\bigr)\,,
\end{alignat}
\end{subequations}
where $v=\vert{d\vec z(t)/dt}\vert$ (the velocity of the particle w.r.t.
the inertial system) and $\gamma(v)=1/\sqrt{1-v^2/c^2}$. This looks all 
very familiar.

\subsection{Energy-momentum conservation in general}
Let's return to general matter models and let $\mathcal{T}^{\mu\nu}$ 
be the total stress-energy tensor of the gravity-matter-system. 
It is the sum of three contributions:
\begin{equation}
T_{\sss\rm total}^{\mu\nu}=
T_{\sss\rm gravity}^{\mu\nu}+
T_{\sss\rm matter}^{\mu\nu}+
T_{\sss\rm interaction}^{\mu\nu}\,,
\label{total-setensor}
\end{equation}
where\footnote{We simply use the standard expression for the 
canonical energy-momentum tensor, which is good enough in the 
present case. If $S=\int L\,dtd^3x$, it is given by 
$T^\mu_\nu:=(\partial L/\partial\Phi_{,\mu})\Phi_{,\nu}-\delta^\mu_\nu L$.}
\begin{subequations}
\label{eq:EM-Tensors}
\begin{alignat}{2}
\label{eq:EM-TensorsGrav}
& T_{\sss\rm gravity}^{\mu\nu}
&&\,=\, \frac{1}{\kappa c^2}\bigl(\partial^{\mu}\Phi\partial^{\nu}\Phi
     -\tfrac{1}{2}\eta^{\mu\nu}\partial_{\lambda}\Phi
      \partial^{\lambda}\Phi\bigr)\,,\\
\label{eq:EM-TensorsMatter}
& T_{\sss\rm matter}^{\mu\nu}
&&\,=\,\text{depending on matter model}\,,\\
\label{eq:EM-TensorsInt}
&T_{\sss\rm interaction}^{\mu\nu}
&&\,=\,\eta^{\mu\nu}(\Phi/c^2) T_{\sss\rm matter}\,.
\end{alignat}
\end{subequations}
Energy-momentum-conservation is expressed by
\begin{equation}
\partial_{\mu}{T_{\sss\rm total}}^{\mu\nu}=F^{\nu}_{\sss\rm external}\,,
\label{em-conservation1}
\end{equation}
where $F^{\nu}_{\sss\rm external}$ is the four-force of a possible
\emph{external} agent. The 0-component of it (i.e. energy conservation)
can be rewritten in the form 
\begin{equation}
\mbox{external power supplied}\ =\
\frac{d}{dt}\int_D d^3x\ {T^{00}_{\sss\rm total}}+
\int_{\partial D}{T^{0k}_{\sss\rm total}}n_k\ d\Omega\,.
\label{em-conservation2} 
\end{equation}
If the matter system is of finite spatial  extent, meaning that outside 
some bounded spatial region $D$ we have 
that $T_{\sss\rm matter}^{\mu\nu}$ vanishes identically, and if we further 
assume that no gravitational radiation escapes to infinity, the  
surface integral in (\ref{em-conservation2}) vanishes identically. 
Integrating (\ref{em-conservation2}) over time we then get 
\begin{equation}
\mbox{external energy supplied}\ =\ \Delta E_{\sss\rm gravity}
+\Delta E_{\sss\rm matter} +\Delta E_{\sss\rm interaction}\,,
\label{em-conservation3}
\end{equation}
with
\begin{equation} 
E_{\sss\rm interaction}=
\int_Dd^3x\ (\Phi/c^2) T_{\sss\rm matter}\,,
\label{interaction-energy}
\end{equation}
and where $\Delta(\mbox{something})$ denotes the difference 
between the initial and final value of $\mbox{`something'}$.
If we apply this to a process that leaves the \emph{internal} 
energies of the gravitational field and the matter system unchanged,
for example a processes where the matter system, or at least the relevant 
parts of it, are \emph{rigidly} moved in the gravitational field, like 
in Einstein's Gedankenexperiment of the `radiation-shaft-system', 
we get
\begin{equation}
\mbox{external energy supplied}\ =\ \Delta\left\{
\int_Dd^3x\ (\Phi/c^2) T_{\sss\rm matter}\right\}\,.
\label{em-conservation4}
\end{equation}
Now my understanding of what a valid claim of energy 
non-conservation would be, is to show that \emph{this} equation can 
be violated, granted the hypotheses under which it was derived. 
This is \emph{not} what Einstein did (compare Conclusions).

If the matter system stretches out to infinity and conducts 
energy and momentum to infinity, than the surface term that was 
neglected above gives a non-zero contribution that must be included 
in (\ref{em-conservation4}). Then a proof of violation of energy 
conservation must disprove this modified equation. 
(Energy conduction to infinity as such is not in any disagreement
with energy conservation; you have to prove that they do not balance
in the form predicted by the theory.) 

\subsection{Conclusion}
For the discussion of Einstein's Gedankenexperiment the term 
(\ref{interaction-energy}) is the relevant one. It accounts for 
the \emph{weight of stress}. Pulling up a radiation-filled box inside 
a shaft also moves up the stresses in the shaft walls that must 
act sideways to balance the radiation pressure. This lifting of 
stresses to higher gravitational potential costs energy, according to 
the theory presented here. This energy was neglected by Einstein,
apparently because it is not associated with a transport of matter. 
He included it in the lowering phase, where the side-walls of the box
are attached to the box and move with it, but neglected them in the 
raising phase, where the side walls are those of the shaft, which 
do not move. But as far as the `weight of stresses' is concerned, 
this difference is irrelevant. What (\ref{interaction-energy}) tells 
us is that raising stresses in an ambient gravitational potential 
costs energy, irrespectively of whether it is associated with an 
actual transport of the stressed matter or not. This would be just 
the same for the transport of heat in a heat conducting material. 
Raising the heat distribution against the gravitational field costs 
energy, even if the material itself does not move. 

I conclude that Einstein's argument is not convincing. Clearly this 
is not meant to give any scientific support to scalar theories of 
gravity (as opposed to GR), which we know are ruled out by experiment.
For example, as already mentioned above, the model theory discussed 
here gives the wrong amount (even the wrong sign) for the perihelion 
shift of Mercury, namely $-1/6$ times Einstein's value. Moreover, 
theories in which the gravitational field couples to matter via its 
trace of the energy-momentum tensor predict a vanishing global 
deflection of light. But what is not the case is that scalar theories 
are \emph{intrinsically} inconsistent, as apparently suggested by 
Einstein. For Einstein this argument might have appeared as a 
convenient physical way to rule out scalar theories, whose 
primary deficiency he saw, however, in the lack of being 
generally covariant.

\subsection*{Acknowledgements}
I thank J\"urgen Ehlers for remarks and discussions.

\end{document}